\documentclass[nofootinbib,preprintnumbers,amsmath,amssymb, preprintnumbers,superscriptaddress,11pt]{revtex4}
\usepackage{amsmath,amssymb,graphicx}
\usepackage[colorlinks=true,citecolor=blue,hyperfootnotes=false]{hyperref}
\usepackage{epsf}
\usepackage{epstopdf}
\usepackage {amssymb}
\usepackage{xcolor}
\newcommand{\nc}{\newcommand}
\nc{\ba}{\begin{eqnarray}}
\nc{\ea}{\end{eqnarray}}
\newcommand\be{\begin{equation}}
\newcommand\ee{\end{equation}}

\newcommand{\calP}{{\cal{P}}}

\newcommand{\calI}{{\cal{I}}}
\newcommand{\xj}{x_{\text{JCSA}}}
\newcommand{\js}{JCSA}
\newcommand{\jss}{JCSA }
\nc{\x}{{\bf{x}}}
\def\nhat{\mathbf{\hat n}}

\def\braket#1#2{\left\langle #1 \left\vert #2 \right.\right\rangle}
\usepackage{array}
\newcounter{magicrownumbers}

\newcommand\rownumber{\stepcounter{magicrownumbers}\arabic{magicrownumbers}}
\newcounter{rowno}
\def\Pfour{{\cal P}_4}
\usepackage[normalem]{ulem}

\begin{document}

\preprint{MIT-CTP-LI/6002}

\author{Alan H. Guth}
\email[Email address: {guth@ctp.mit.edu}]{}
\affiliation{Department of Physics, Laboratory for Nuclear
Science, and Center for Theoretical Physics -- A Leinweber
Institute, Massachusetts Institute of Technology, Cambridge, MA
02139}

\author{Mohammad Hossein Namjoo}
\email[Email address: {mh.namjoo@ipm.ir}]{}
\affiliation{School of Astronomy, Institute for Research in Fundamental Sciences (IPM), Tehran,
Iran}

\date{February 16, 2026}
\title{Statistical isotropy of the universe and the look-elsewhere effect}

\begin{abstract}

Recently, Jones et al.~\cite{Jones:2023ncn} claimed strong
evidence for the statistical anisotropy of the universe. The
claim is based on a joint analysis of four different anomaly
tests of the cosmic microwave background data, each of which is
known to be anomalous, with a lower level of significance. They
reported a combined $p$-value of about $3\times 10^{-8}$, which
is more than a $5\sigma$ level of significance. We observe that
statistical anisotropy is not even relevant for two of the four
considered tests, which seems sufficient to invalidate the
authors' claim.  Furthermore, even if one reinterprets the claim
as evidence against $\Lambda$CDM rather than statistical
anisotropy, we argue that this result significantly suffers from
the look-elsewhere effect. Assuming a set of independent (i.e.,
uncorrelated) tests, we show that if the four tests with the
smallest $p$-values are cherry-picked from 10 independent tests,
the $p$-value reported by Jones et al. corresponds to only
$3\sigma$ significance. If there are 27 independent tests, the
significance falls to $2\sigma$. These numbers, however,
overstate our argument, since the four tests used by Jones
et al. are slightly correlated. Determining the correlation of
Jones et al.'s tests by comparing their joint $p$-value with the
product of the four separate $p$-values, we find that about 16 or
50 tests are sufficient to reduce the significance of Jones et
al.'s results to 3$\sigma$ or 2$\sigma$ significance,
respectively. We also provide a list of anomaly tests discussed
in the literature (and propose a few generalizations), suggesting
that very plausibly 16 (or even 50) independent tests have been
published, and possibly many more have been considered but not
published. We conclude that the current data is consistent with
the $\Lambda$CDM model and, in particular, with statistical
isotropy.

\end{abstract}

\maketitle
\newpage

\def\refname{{\hskip -22.27782pt}References}
\def\acknowledgmentsname{{\hskip -22.27782pt}Acknowledgments}

\tableofcontents
\section{Introduction}
\label{sec:intro}
 
The $\Lambda$CDM model has demonstrated remarkable success in describing cosmological observations. Nevertheless, any deviation from $\Lambda$CDM holds significant interest as it may indicate the existence of new physics. A large number of anomaly tests have been extensively investigated in the literature, some of which exhibit discrepancies from the $\Lambda$CDM predictions, particularly at large cosmological scales. However, so far, no single test  has found a sufficiently significant deviation to cause $\Lambda$CDM to be rejected by the community.

Recently, Jones, Copi, Starkman and Akrami~\cite{Jones:2023ncn}
(\js) combined four different anomaly tests and, by a joint
analysis, claimed the rejection of statistical anisotropy by more
than $5\sigma$ significance. The four tests considered in \jss
are those that have already been known to signal some (but not
very significant) deviation from $\Lambda$CDM.  Specifically, the
considered tests are {(i) the low-level of large-angle cosmic
microwave background (CMB) temperature correlations,\footnote{It
is worth noting that while Planck, along with JCSA, used a 26\%
masked CMB map, recently Ref.~\cite{Herold:2025mro} analyzed this
test with a CMB map that is only 1\% masked.
Ref.~\cite{Herold:2025mro} finds that the significance of the
low-level of large-angle CMB correlation decreases from about
$3\sigma$ to about $2\sigma$. This new finding is not directly
relevant to this paper, so we will discuss it no further.}  (ii)
the excess power in odd versus even low-$\ell$ CMB multipoles,
(iii) the low variance of large-scale CMB temperature
anisotropies in the ecliptic north (compared to the south), and
(iv) the alignment and planarity of the quadrupole and octopole
of the CMB temperature anisotropies}. 

We note that the first two tests actually measure deviations from
$\Lambda$CDM but not statistical anisotropy. Any assignment of
values for the $C_\ell$'s is consistent with statistical
isotropy. This alone seems to invalidate the authors' claim ---
as stated in their title --- to have shown that ``the universe is
not statistically isotropic." However, one can still consider the
analysis of \jss as a claim that a statistically significant
deviation from $\Lambda$CDM has been detected.  In this paper, we
argue that this claim suffers significantly from the
look-elsewhere effect.  \jss recognized that look-elsewhere
effects were relevant, but they assumed that their results were
so strong that look-elsewhere effects could not possibly call
them into question: ``While there are undoubtedly look-elsewhere
penalties to be paid for this collection of mostly {\it a
posteori} statistical anomalies, it is clear that there is very
strong evidence in the CMB for the violation of statistical
isotropy.'' Here we argue that their results are in fact
undermined by the look-elsewhere effect.  We conclude that the
current data is still consistent with $\Lambda$CDM (and, in
particular, with a statistically isotropic universe).

Even before the phrase ``look-elsewhere effect" began to appear
frequently in the scientific literature, scientists have been
aware that neglecting this effect can lead to false conclusions.
A good example in astronomy is the effort by Halton Arp, starting
in the 1960s, to challenge the foundational assumption that
redshift is a reliable indicator of distance.  In evaluating this
history, we should keep in mind that Arp was a prominent
astronomer, with a Ph.D. from the California Institute of
Technology, who was recognized by the Helen B. Warner Prize of
the American Astronomical Society and the Newcomb Cleveland Prize
of the American Association for the Advancement of Science.  Arp
was a staff member at the Mount Wilson and Palomar Observatories
for 29 years, starting in 1957. 
% Sources: dates from Wikipedia, name of organization from Martin
%   Rees review in Physics Today of ``Quasars, Redshifts and
%   Controversies''
Arp's arguments were based mainly on finding collections of two
or more objects with significantly different redshifts, but which
have features that, according to Arp's claims, would be highly
unlikely unless the objects had some physical connection, which
would require them to be at about the same distance.  But
look-elsewhere effects were typically ignored.  For example, in
Ref.~\cite{arp-1967}, Arp argued that two radio sources that are
near galaxy No.~145 in his {\it Atlas of Peculiar Galaxies} have
locations that indicate that they were almost certainly ejected
from the galaxy.  Based on the angular distance of the radio
sources from the galaxy, and the angular distance of the galaxy
from the midpoint of the two radio sources, Arp calculated that
the probability of finding such a triplet ``at an arbitrary point
in the sky'' is only one in $4 \times 10^5$.  He stopped there,
never considering the probability of finding such a configuration
{\it somewhere} in the sky!

In December 1972 there was a well-publicized debate
\cite{field1973redshift} on these issues between Arp and John N.
Bahcall at a meeting of the American Association for the
Advancement of Science.  In rebutting Arp's case, Bahcall stated
a clear ``moral'':
\begin{quote}
%	The moral of the story of the
%	apparent connection between NGC 4319 and M 205 is clear: 
Seek and ye shall find, but beware
	of what you find if you have to work very hard to see something
	you wanted to find.
\end{quote}
Arp's proposal of anomalous redshifts was never generally
accepted, although he made his point of view widely known, and he
was supported by a few leading astronomers, such as Geoffrey 
Burbidge 
and Margaret Burbidge.  In 1983 the telescope allocation
committee at Palomar sent Arp a letter stating, as summarized by
Arp in Ref.~\cite{arp-1987}, that Arp's ``research was judged to
be without value and that they intended to refuse allocation of
further observing time.''  In his review of Ref.~\cite{arp-1987}
in {\it Physics Today} in 1988 \cite{rees-1988}, Martin Rees
wrote ``Most astronomers who have followed Arp's work over the
years have judged that his case for anomalous redshifts lacks
cumulative weight, and has even weakened as extragalactic
astronomy has advanced.''  Today there seems to be very little if
any support in the astronomical community for Arp's views.

In this paper, we show that the look-elsewhere effect significantly weakens the analysis of \js, even when we set aside the issue that two out of the four considered tests do not test statistical anisotropy. In Sec.~\ref{sec:dist-calc}, we calculate the probability distribution of the joint $p$-value of the four most discrepant tests when a larger set of independent tests is considered. By studying the properties of this distribution, we show that the significance of the result reported in \jss reduces to $3\sigma$ if these four tests are cherry-picked from 10 independent tests, or even to $2\sigma$ if they are picked from 27 tests. In Sec.~\ref{sec:correlations}, we use a simple method to roughly account for the correlation among the tests considered in \js,  and estimate that the number of independent tests needed to reduce the significance to $3\sigma$ or $2\sigma$ increases to 16 or 50, respectively. In Sec.~\ref{sec:tests}, we list a number of different anomaly tests that have appeared in the literature and  propose a few generalizations, suggesting that 16 to 50 tests have plausibly been performed. 

\section{Statistics of the product of the four smallest $p$-values}
\label{sec:dist-calc}

In this section we obtain the probability distribution  for the
product of the four most anomalous (i.e., smallest) $p$-values
among the $p$-values associated with a number of independent
tests, under the assumption that there are no anomalies --- i.e.,
under the assumption that the outcomes of the tests obey exactly
the probability distribution assumed in the calculation of the
$p$-values.  We denote the total number of tests by $n_T$, the
$p$-values by $p_i$ (for $i = 1,\dots,n_T$), the product of the
four smallest $p$-values by $x$, and its probability density by
$\Pfour(x)$.  

To simplify the analysis, we first consider the situation where
$p_1 < p_2 < \dots < p_{n_T}$. In this case, we have $x = p_1 p_2
p_3 p_4$, which simplifies the calculation of the probability
distribution.  The probability density for $x$, given that $p_1 <
p_2 < \dots < p_{n_T}$, can be written as
\be
\Pfour(x|p_1 < p_2 < \dots < p_{n_T}) = {P(x \hbox{ and } p_1
	< p_2 < \dots < p_{n_T}) \over 
	P(p_1 < p_2 < \dots < p_{n_T}) }\ ,
\ee
where $P(\tilde x \hbox{ and } p_1 < p_2 < \dots < p_{n_T}) d x$
is the probability that $x$ lies between $\tilde x$ and $\tilde x
+ d x$, and that $p_1 < p_2 < \dots < p_{n_T}$.  Since there are
$n_T$!  equally likely permutations of $p_1, \dots, p_{n_T}$, we
have $P(p_1 < p_2 < \dots < p_{n_T}) = 1/n_T!$.  

Note that the $p$-value, assuming  that there are no
anomalies, has a uniform distribution, regardless of the
distribution of the measured random variable. This is because, by
definition, the probability that the $p$-value is less than $x$
is always equal to $x$.  Thus, since the $p_i$'s obey uniform
distributions, we can write ${\cal I}(x) \equiv P(x \hbox{ and } p_1
< p_2 < \dots < p_{n_T})$ as
\ba 
\calI(x)= \int_0^1 \prod_{i=1}^{n_T} dp_i \, \delta( p_1 p_2 p_3 p_4 -x) \, \theta(p_2-p_1, p_3-p_2,..., p_{n_T}-p_{n_{T-1}}) \ ,
\ea 
where $\theta$ is a generalized Heaviside theta function that is 1 if all of its arguments are positive and 0 otherwise. Performing the integrals  over the variables $p_5$ to $p_{n_T}$  is simple due to the irrelevance of the delta function. We have 
\ba
\label{high_i_int}
  \int_0^1 \prod_{i=5}^{n_T} dp_i \, \theta(p_5-p_4, p_6-p_5,..., p_{n_T}-p_{n_{T-1}}) = \dfrac{1}{n_d!}\, (1-p_4)^{n_d}\ ,
\ea 
where $n_d \equiv n_T-4$. To justify this relation, note that the
$\theta$-functions require, for all $i$ in the range 5 to $n_T$,
that $p_4 < p_i < 1$ and that the $p_i$ are ordered.  But the
ordering does not affect the integral. If the ordering
requirement were dropped, all $n_d!$ orderings would contribute
equally, and the region of integration would be a cube of volume
$(1-p_4)^{n_d}$.  Thus, the integration with the ordering
requirement is given by Eq.~(\ref{high_i_int}).

Thus we are left with four remaining integrals over $p_1,...,p_4$.  We start with $p_1$, which removes the delta function (and changes the argument of the theta function), and then successively perform the integrals over  $p_2$, $p_3$, and  $p_4$ as follows
\ba
\calI(x) &=& \dfrac{1}{n_d!}\int_0^1 \prod_{i=2}^{4} dp_i \, \dfrac{(1-p_4)^{n_d}}{p_2p_3p_4}\, \theta \bigg(p_2- {\left[\frac{x}{p_3p_4}\right]^{1/2}},  p_3-p_2,p_4-p_3\bigg) \ ,
\\
&=& \dfrac{1}{n_d!}\int_0^1   dp_3 dp_4 \, \ln  {\left(\dfrac{p_3^3 \, p_4}{x}  \right)}\dfrac{(1-p_4)^{n_d}}{2p_3p_4}\, \theta \bigg(p_3-\left[\frac{x}{p_4}\right]^{1/3},p_4-p_3 \bigg) 
\\ 
&=&   \dfrac{1}{n_d!}\int_0^1   dp_4  \dfrac{(1-p_4)^{n_d}}{12p_4} \ln^2 \left(\dfrac{p_4^4}{x}  \right) \, \theta \big(p_4-x^{1/4}\big) 
\\
&=&  \dfrac{1}{12n_d!} {\cal J}_{n_d} (x) ,
\ea 
where 
 \ba 
{\cal J}_{n_d}(x) &\equiv &   \int_{x^{1/4}}^1   dp_4  \dfrac{(1-p_4)^{n_d}}{p_4} \ln^2 \left(\dfrac{p_4^4}{x}  \right) \label{p4int}\\
&=&    -  \sum_{m=0}^{n_d} \binom{n_d}{m} \int_{x^{1/4}}^1   dp_4   (-p_4)^{m-1} \ln^2 \left(\dfrac{p_4^4}{x}  \right)
\\
&=& {-\dfrac{1}{12}\ln^3 x+ \sum_{m=1}^{n_d} \frac{(-1)^{m}}{m^3}\binom{n_d}{m}
\left[ m (m \ln x +8) \ln x 
+32 (1-x^{m/4})
\right]}
\\
&=&-\frac{1}{12} \big(4 H_{n_d}+\ln x\big)^3 
-\frac{2}{3} \big(4 H_{n_d}+\ln x\big) \big(\pi ^2-6 \psi ^{{(1)}} (n_d+1)\big) -\frac{32}{3}  \zeta (3) 
\nonumber \\
&&+32 n_d \, x^{1/4} \, _5F_4\left(1,1,1,1,1-n_d;2,2,2,2;x^{1/4}\right)- \frac{16}{3} \psi ^{(2)}(n_d+1)\ ,\label{analytic}
\ea 
where $H_{n_d}$ is the $n_d$-th harmonic number, $\psi ^{(i)}(.)$
is the polygamma function of order $i$, $_5F_4(.)$ is the
hypergeometric function, and $\zeta(.)$ is the Riemann zeta
function \cite{wolfram}.  Note that numerical integration of
Eq.~(\ref{p4int}) is an effective method of determining
$\calP_4(x)$, but Eq.~(\ref{analytic}) allows the answer to be
expressed in terms of named functions.  The probability density
for $x$ would not change if the $p$-values $p_1,...,p_{n_T}$
occurred in a different order, so
\ba 
\label{P4_final}
\calP_4(x) =\Pfour(x|p_1 < p_2 < \dots < p_{n_T}) =   \dfrac{n_T!}{12n_d!} {\cal J}_{n_d}(x).
\ea 

\begin{figure}
	\includegraphics[scale=1]{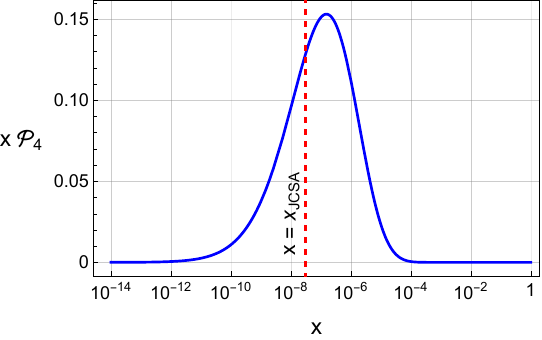}
	\caption{Plot of $x \calP_4(x)$ for $n_T=100$. $x
     \calP_4(x)$ is the probability density for $\ln x$.  Since
     $x$ is shown on a logarithmic scale, the area under this
     curve is proportional to the probability that $x$
     lies in a given range.  The red vertical dashed line shows
     $x = \xj$, the value of $x$ found by \js, which can be seen
     to be in the region of high probability.  Numerical
     integration shows that for this case, there is a probability
     of 33.8\% that $x$ will be smaller than $\xj$.}
	\label{fig:pdf100}
\end{figure}

 We can now study different properties of this  distribution. Fig.~\ref{fig:pdf100} depicts the behavior of $\calP_4(x)$ for $n_T=100$. Clearly, $\calP_4(x)$ is highly {asymmetric. In fact, it diverges like $\ln^3 x$ as $x\to 0$, while it approaches $0$ as $x\to 1$. For this distribution the mean is significantly larger than the median.} While, by definition, there is 50\% chance that  a randomly drawn sample is larger than the median, it is less likely to be larger than the mean. In this sense, the mean is atypical. As a specific example, for $n_T=100$ the probability that the measured value is larger than the mean is only {about} 17\%. We conclude that the median is a more appropriate  {measure of central tendency of the probability distribution.}\footnote{{The median is generally preferred by statisticians as a measure of central tendency, especially for skewed distributions, as stated for example by the Australian Bureau of Statistics  \cite{Aust_Bureau}: `` The median is less affected by outliers and skewed data than the mean and is usually the preferred measure of central tendency when the distribution is not symmetrical." }}

 \begin{figure}
	\includegraphics[scale=1]{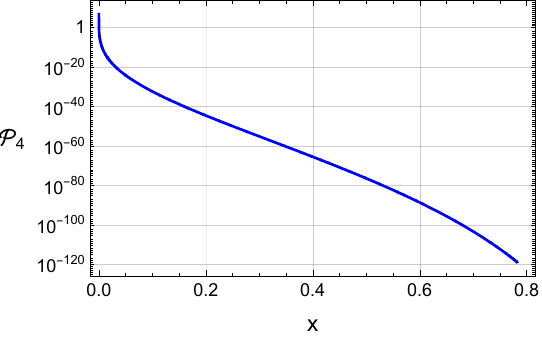}
	\caption{The probability distribution $\calP_4(x)$ for $n_T=100$. }
	\label{fig:pdf100-v1}
\end{figure}

While Fig.~\ref{fig:pdf100-v1} illustrates well the
skewness of $\calP_4(x)$, it makes it hard to see what is likely,
since almost all of the probability is concentrated in the $\ln^3 x$
divergence of $\calP_4(x)$ at $x=0$.  To better understand the
likely outcomes, it is more useful to plot $x \calP_4(x)$, 
the probability density for $\ln x$, which is shown in
Fig.~\ref{fig:pdf100}.  The graph shows clearly that, for
$n_T=100$, the value of $x$ found by \jss is quite probable.

\begin{figure}
 	\includegraphics[scale=1]{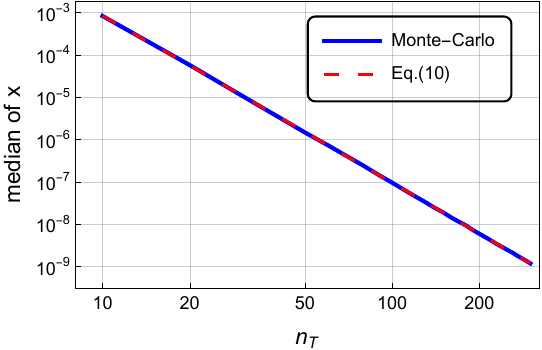}
 	\caption{The median of $\calP_4$ as a function of $n_T$ and its comparison with the result of the Monte Carlo simulation. For $n_T=133$, we obtain $x \simeq  \xj$ as the median of $\calP_4(x)$.}
 	\label{fig:median}
 \end{figure}

One can ask how many independent tests are needed so that the median of the distribution is equal to the $p$-value  reported in \js, i.e., $x_{\text{JCSA}}=3\times 10^{-8}$.  Fig.~\ref{fig:median} depicts the behavior of the median of $\calP_4$ (calculated numerically {from Eq.~\eqref{P4_final}}) as a function of $n_T$.    {One can see that if $133$ tests are considered, $x$ would be less than $\xj=3\times 10^{-8}$ about half the time. }
 
As a check, in Fig.~\ref{fig:median}, we also compare the
aforementioned semi-analytic result with the result of
Monte-Carlo simulations assuming a Gaussian distribution with
zero mean and standard deviation 1 for each test. For each
$n_T=10, 20, ... , 300$, we generated $10^5$ random samples of
$(p_1,p_2,...,p_{n_T})$, calculating $x$ (the product of the four
smallest $p$-values). We plot the median of these $x$ values for
each $n_T$, joining the points to get a smooth line.  As can be
seen, the Monte Carlo trials agree beautifully with the
calculations based on our calculation of $\calP_4(x)$.
 
  \begin{figure}
 	\includegraphics[scale=1]{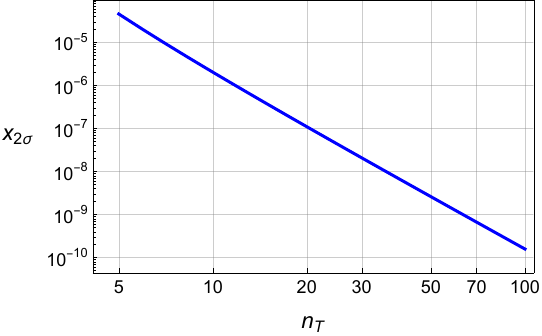}
 	\caption{ { The measured value of $x$ with $2\sigma$ significance, $x_{2\sigma}$, according to the probability distribution $\calP_4$, as a function of $n_T$. The $2\sigma$ significance of  $ x=  \xj$ requires $n_T \simeq  27$.} }
 	\label{fig:xD}
 \end{figure}
 
  \begin{figure}
 	\includegraphics[scale=1]{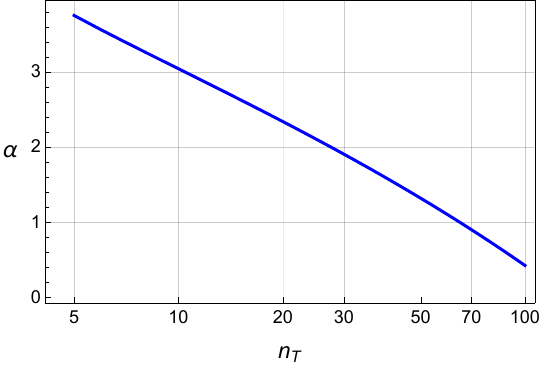}
 	\caption{ The {statistical} significance $\alpha$ (in units of $\sigma$) when the measured value is $x = \xj$,  as a function of $n_T$, according to the probability distribution $\calP_4$. The measurement of $x=  \xj$  corresponds to roughly $2\sigma$ significance if  $n_T = 27$ and to $3\sigma$ significance if  $n_T = 10$. }
 	\label{fig:alpha}
 \end{figure}

Insisting that $\xj$ be the median of the distribution is perhaps
a stronger requirement than needed to discredit the claims of
statistical anisotropy. We may instead ask how many independent
tests are needed so that $\xj=3\times 10^{-8}$ corresponds to the
significance of, say, $2\sigma$ (97.72\% CL) or $3\sigma$
(99.86\% CL).  ({By contrast, \jss claim a more than $5\sigma$
level of significance}.) Fig.~\ref{fig:xD} shows the behavior of
the measured value {of $x$ with $2\sigma$ significance, which we
denote by $x_{2\sigma}$,} as a function of $n_T$. Also, in
Fig.~\ref{fig:alpha} we show how the statistical significance
$\alpha$ of $x=\xj$, measured in units of $\sigma$, varies with
$n_T$.  We see that $x=\xj$ corresponds to $2\sigma$ significance
if $n_T \simeq 27$ and to $3\sigma$ significance if $n_T \simeq
10$.  The possibility of $n_T=10$ or $27$ tests seems plausible
enough to call into serious question the claims of
\js\null.  In Sec.~\ref{sec:tests}, we provide a list of
different anomaly tests that are already discussed in the
literature (and also propose some generalizations).

\section{{Accounting for correlations}}
\label{sec:correlations}

So far, we have assumed that all  tests are independent. However, the four tests considered in \jss are not completely uncorrelated. \jss define the ``correlation factor", which we denote by $\cal C$, by
\ba 
{\cal C }\equiv  \dfrac{\xj}{p_1 \, p_2 \, p_3 \, p_4}\,  \simeq 51\, ,
\ea  
where the $p_i$ are the individual $p$-values for the tests considered. 

{To approximately account for this correlation in a tractable way, we first estimate the ``effective number of independent tests", $n_{\rm{eff}}$, as follows. We let $\bar p \equiv (p_1 \, p_2 \, p_3 \, p_4)^{1/4}$ be the geometric mean of the $p$-values, and define $n_{\text{eff}}$ via $\bar p^{\, n_{\text{eff}}}\equiv \xj$.}
For the numbers reported by \js, we obtain
\ba 
n_{\rm{eff}} \simeq 3.26.
\ea 
As expected, $n_{\rm{eff}}<4$ due to correlations. 

\begin{figure}
	\includegraphics[scale=1]{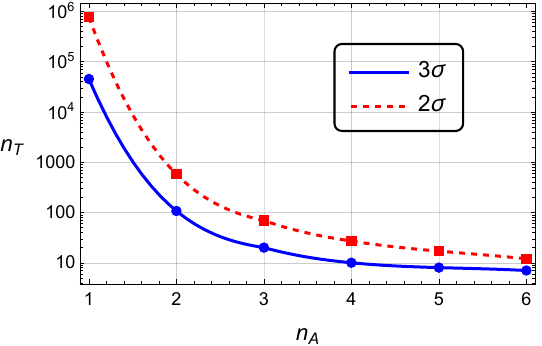}
	\caption{The data points show the number of independent tests $n_T$ required to reduce $\xj$ to very nearly 2$\sigma$ or 3$\sigma$ as a function of $n_A$, the number of independent tests that are combined to determine the joint $p$-value. The  interpolating curves are used to generalize this result to define $n_T$ as a function of the effective (non-integer) $n_A$. For the data in \js, $n_A=3.26$.}
		\label{fig:nAnT}
\end{figure}

Next, we generalize the calculations of Sec.~\ref{sec:dist-calc},
considering the combination of the $n_A$ {most discrepant tests
(rather than the 4 most discrepant tests)} out of a set of $n_T$
independent tests.  We leave the details to
Appendix~\ref{sec:App}, where we find that $\calP_{n_A}(x)$ can
be written as
\be
\label{PnA}
\calP_{n_A}(x) = 
\begin{cases}
n_T (1-x)^{n_D} & \hbox{if $n_A = 1$} \\
\displaystyle
\frac{n_T!}{n_D!} \frac{n_A^2 (n_A-1)}{(n_A!)^2}
\int_{x^{1/n_A}}^1 d p \, \frac{(1-p)^{n_D}}{p} \ln^{n_A-2} \left(
\frac{p^{n_A}}{x} \right) \ . & \hbox{otherwise}\, ,\\
\end{cases}
\ee
where $n_D\equiv n_T-n_A$.
For each value of $n_A=1,..,6$, we find the integer value of
$n_T$ that comes closest to reducing the significance of $\xj$ to
2$\sigma$ or 3$\sigma$. Fig.~\ref{fig:nAnT} shows a plot of $n_T$
versus $n_A$, along with interpolating curves.\footnote{We
applied Mathematica's built-in ``Interpolation" (a third order
spline with ``not-a-knot" boundary conditions) to a table of
pairs $(n_A, \ln n_T)$.} Using these curves to determine $n_T$
for $n_A=3.26$, we find that we need $n_T\simeq 50$ or $n_T\simeq
16$ tests to reduce $\xj$ to 2$\sigma$ or 3$\sigma$ significance,
respectively. 
	
\section{A list of anomaly tests and possible generalizations}
\label{sec:tests}
\begin{table}	
	\caption{A list of anomaly tests that have appeared in the literature. %Comments: The Q-O-alignment  is a subset of the cross-cross MVs test. 
	}
	\centering
	\begin{tabular}{|c|c|c|c|}%c|c|}
		\hline
	&	Test & Parameters &Ref.   \\
		%Symbol & variables &Ref.\\
		\hline  \hline  
	\rownumber &	quadrupole-octopole alignment &  --- & \cite{Planck:2013lks} \\
		\hline 
	\rownumber &		hemispherical asymmetry &  $\ell_{\max}$  & \cite{Planck:2013lks} \\
		\hline 
%	\rownumber &		phase correlations (scaling index)&  $r$ (scale parameter)  & \cite{Planck:2013lks} \\
%		\hline 
%	\rownumber &		 power asymmetry & --- & \cite{Planck:2013lks}  \\
%		 \hline 
	\rownumber &	local variance asymmetry & disc radius &\cite{Akrami:2014eta}  \\
		 \hline 
	\rownumber &	generalized modulation	
	% Bipolar spherical harmonics 
	 & L(type of modulation) \& $\ell_{\text{bins}}$ & \cite{Planck:2013lks} \\
		 \hline 
		 	\rownumber &		vector-vector of multipole vectors (MVs)& $\ell_1$ \& $\ell_2$ & \cite{Copi:2003kt}\\
		 \hline 
		 	\rownumber &		 vector-cross of MVs & $\ell_1$ \& $\ell_2$ &\cite{Copi:2003kt} \\
		 \hline 
		 	\rownumber &		 cross-cross of MVs & $\ell_1$ \& $\ell_2$  &\cite{Copi:2003kt} \\
		 \hline 
		 	\rownumber &		 oriented area & $\ell_1$ \& $\ell_2$ &\cite{Copi:2003kt} \\
		 \hline 
 	 	  	\rownumber & histograms  of MV angular distribution &   $\ell_{\min}$  \& $\ell_{\max}$ \& bin-size& \cite{Oliveira:2018sef}\\
    	 	\hline 
  	\rownumber &  histograms  of Fr\' echet vector  angular distribution & $\ell_{\min}$  \& $\ell_{\max}$ \& bin-size & \cite{Oliveira:2018sef}\\
	\hline 
\rownumber &		 mirror parity & $N_{\text{side}}$ &\cite{Planck:2013lks}  \\
\hline 
  	\rownumber &		  cold spot & $R$ (scale) \& $\nu$ (threshold) & \cite{Planck:2013lks} \\
\hline 
	\rownumber &		point parity asymmetry & $\ell_{\max}$ &\cite{Planck:2013lks}  \\
\hline 
 	\rownumber & entropy  &$\ell$ & \cite{Helling:2006xh,Minkov:2018prz}\\
  	\hline 	
  	\rownumber &	 variance, skewness, kurtosis &  $N_{\text{side}}$ & \cite{Planck:2013lks} \\
  	\hline 
  	\rownumber &	 large-angle correlation &  $\theta_{\min}$ & \cite{Planck:2013lks} \\
  	\hline 
   	 	\rownumber & bispectrum, trispectrum & scale and configuration dependence & \cite{Planck:2019kim} \\
    	 	\hline 
	\end{tabular}
	\label{Tab:tests}
\end{table}

Our analysis of 
Sec.~\ref{sec:correlations}
demonstrates that having 16 to 50 independent tests suffices to
substantially reduce the significance of $\xj$. In
Table~\ref{Tab:tests}, we provide examples of anomaly tests that
have appeared in the literature. The first 12 tests measure
deviations from statistical anisotropy while the last 5 tests
measure other deviations from $\Lambda$CDM\null. The tests are not
necessarily independent. However, we believe that the list is
long enough to make the existence of the required number of
independent tests very plausible. {In addition,} each test in
Table~\ref{Tab:tests} contains free parameters, different values
of which may also be considered as different tests.  It is worth
noting that, under $\Lambda$CDM, different CMB multipoles
$a_{\ell m}$ are independent in the sense that the joint
probability density factorizes, $p(a_{\ell m} , a_{\ell'
m'})=p(a_{\ell m}) \, p(a_{\ell' m'})$ for $\ell \neq \ell'$ or
$m\neq m'$.  Thus, any statistic which involves a range of
$\ell$'s could produce a number of independent tests by applying
the statistic to different, non-overlapping ranges of $\ell$'s.

We also stress that the number of tests that have actually been performed is unknown, and may be much larger than the number reported in the literature due to publication bias. That is, tests which  search for inconsistencies with  $\Lambda$CDM, but do not find any, are likely to remain unpublished.

To explore the possibilities for tests beyond those in Table
\ref{Tab:tests}, we briefly mention a few plausible
generalizations.  For example, a very general suite of modulation
tests can be constructed using the spherical harmonic correlation
matrix, defined by
\ba A^{LM}_{\ell_1 \ell_2} \equiv \sum_{m_1 m_2} a_{\ell_1 m_1}
a_{\ell_2 m_2} \braket{\ell_1 m_1 \ell_2 m_2}{LM} \ ,
\ea 
where the $\braket{\ell_1 m_1 \ell_2 m_2}{LM}$ are
Clebsch-Gordon coefficients (see Ref.~\cite{Planck:2013lks} for further
discussion).  These coefficients completely describe the
two-point function, in the sense that the integral
\ba 
\int d \Omega_{\nhat} \, d \Omega_{\nhat'} \, w(\nhat,\nhat')
\, \delta T(\nhat) \, \delta T(\nhat') \ ,
\ea 
for any weight function $w(\nhat, \nhat')$, can be expressed as a
linear sum of $A^{LM}_{\ell_1 \ell_2}$'s.

Using these coefficients, tests are proposed by a weighted sum of
functions of $A^{LM}_{\ell_1 \ell_2}$
\cite{Planck:2013lks,Souradeep:2006dz}.\footnote{{For example,
Ref.~\cite{Souradeep:2006dz} studies the variable $\kappa_L =
\sum_{\ell_1,\ell_2,M} W_{\ell_1} W_{\ell_2} |A_{\ell_1
\ell_2}^{LM}|^2$ where $W_{\ell}$ is a window function that
smooths the map in real space.}} However, each coefficient
$A^{LM}_{\ell_1 \ell_2}$, for any value of $\ell_1 > 1$, $\ell_2
> 1$, and $L > 0$, can also be considered as a different test of
anisotropy.  (For $L=0$, nonzero values of $A^{00}_{\ell \ell}$
are consistent with isotropy.) There is good motivation, however,
to always sum over $M$, as $\sum_M |A^{LM}_{\ell_1 \ell_2}|^2$,
to avoid quantities that depend on our arbitrary choice of
coordinate axes.\footnote{Unlike $A^{LM}_{\ell_1 \ell_2}$ for
$L\neq 0$, quantities like $\sum_M |A^{LM}_{\ell_1 \ell_2}|^2$
are expected to be nonzero even in the case of statistical
isotropy. Therefore, it is appropriate to study the ``unbiased"
test by subtracting the expected value according to $\Lambda$CDM,
which can be obtained as a function of $C_\ell$'s
\cite{Souradeep:2006dz}.}

As another example, one can define a test that measures the
periodicity-in-$\ell$ of the CMB power spectrum, as a
generalization of the {point} parity asymmetry test.  To be more
explicit, note that the parity asymmetry test compares the even
and odd parity components of the CMB power spectrum, which are
defined by \cite{Planck:2013lks,Jones:2023ncn}:
\ba 
\label{D+-}
D_{\pm}  =\dfrac{2}{\ell_{\max}-1} \sum_{ \ell =2}^{\ell_{\max}}\dfrac{\ell (\ell+1)}{2\pi} \dfrac{\big(1\pm (-1)^{\ell} \big)}{2}  C_\ell\, .
\ea 
Note that $D_{+}$ and $D_{-}$ are weighted sums of $C_\ell$'s. {In principle, the point parity asymmetry test could be generalized to consider $C_\ell$ for each $\ell$ as a distinct test.}  However, to minimize the statistical noise, it is beneficial to combine a set of $C_\ell$'s, as is done, for example, in Eq.~\eqref{D+-}. A class of generalizations of the parity asymmetry test would be to compare weighted sums of $C_\ell$'s grouped by {$\ell\, \, \text{modulo}\, \, p$} for any integer $p$.  The  range of included $\ell$'s can also be varied, as well as the   weighting scheme.

\section{Conclusion}
We have shown that the claim of observed statistical anisotropy by \jss is flawed in at least two ways. First, we noted that two of the four tests considered by \jss do not actually test  statistical anisotropy, although they are tests of $\Lambda$CDM.
	
Second, even if the \jss result is reinterpreted as evidence against $\Lambda$CDM, we showed that the result significantly suffers from the look-elsewhere effect. To explore the look-elsewhere effect, we calculated the probability distribution for the combined $p$-value of {the} $4$ most discrepant tests out of a total of $n_T$ independent tests.  Assuming that the four tests considered by \jss  are independent, we found that the significance of  {the} \jss result is reduced to  $3\sigma$  if the four tests are cherry-picked from 10 independent tests. If the tests are picked from 27 independent tests, the significance is reduced to $2\sigma$. By roughly accounting for the correlation among the tests considered in \js,  we estimated that these numbers increase to 16 or 50 independent tests, respectively. To explore tests that have been reported, we have constructed a  list of 17 tests that have appeared in the literature (Table~\ref{Tab:tests}).  Many of these tests involve choices of parameters, offering the possibility of multiple tests by making different choices. We also argued that the number of tests that have actually been performed may be much larger than the number reported in the literature due to publication bias --- that is, tests which find no tension with $\Lambda$CDM may remain unreported. To explore the possibilities for tests beyond those in Table~\ref{Tab:tests}, we suggested a few plausible generalizations. 

We conclude that 16 to 50 independent tests of $\Lambda$CDM have plausibly been carried out, and therefore the current data is consistent with $\Lambda$CDM and, in particular, with the statistical isotropy of the universe. 

\acknowledgments
We thank  G. Efstathiou for useful discussions. 
A.H.G.'s work was supported in part by
the U.S. Department of Energy, Office of Science, Office of High
Energy Physics under grant Contract Number DE-SC0012567.

\addcontentsline{toc}{section}{Appendices:}

\appendix
\section{Proof of Eq.~(\ref{PnA})}
\label{sec:App}

Generalizing slightly the arguments in the main text, the
probability density $\calP_{n_A}(x)$ for $x=p_1 p_2 \ldots
p_{n_A}$, where $p_1, p_2, \ldots p_{n_A}$ are the $n_A$ smallest
$p$-values out of a set of $n_T$ $p$-values, is given by
\be
\calP_{n_A}(x) = n_T! \int_0^1 \prod_{i=1}^{n_T} \, d p_i \,
     \delta(p_1 p_2 \ldots p_{n_A} - x) \, \theta(p_2-p_1,
     p_3 - p_2,  \, \ldots, \, p_{n_T} - p_{n_T-1})
     \ .
     \label{PnA2}
\ee
As discussed in the text, the integrals over $p_{n_A+1}$ ...
$p_{n_T}$ can be carried out immediately: 
\be
  \int_0^1 \prod_{i=n_A+1}^{n_T} \, d p_i \, \theta(p_{n_A+1} -
     p_{n_A}, \, \ldots, \, p_{n_T} - p_{n_T-1}) = {1 \over
     n_D!} \, (1 - p_{n_A})^{n_D} \ ,
     \label{intnD}
\ee
where
\be
  n_D \equiv n_T - n_A \ .
\ee

For $n_A=1$ or 2, these equations lead immediately to
\be
\begin{aligned}
  \calP_{1}(x) &= n_T (1 - x)^{n_D} \ ,\\
     \calP_{2}(x) &= {n_T! \over n_D!} \int_{x^{1/2}}^1 {d p_2
     \over p_2} (1 - p_2)^{n_D} \ ,
\end{aligned}
\ee
both of which agree with Eq.~(\ref{PnA}) in the text.
For $n_A > 2$, we can combine Eqs.~(\ref{PnA2}) and (\ref{intnD})
and integrate over $p_1$, using the $\delta$-function.  We then have
\be
\begin{aligned}
\calP_{n_A}(x) &= {n_T! \over n_D!} \int_0^1 \prod_{i=2}^{n_A}
     \, d p_i \, {(1 - p_{n_A})^{n_D} \over p_2 \ldots p_{n_A} }
     \, \theta\left(p_2-{x \over p_2 p_3 \ldots
     p_{n_A}}, \, p_3 - p_2, \,
     \ldots, \, p_{n_A} - p_{n_A-1}\right)  \\
   &= {n_T! \over n_D!} \int_0^1 {d p_{n_A} \over p_{n_A}} \,
     (1-p_{n_A})^{n_D} \, F_{n_A} (p_{n_A}, x/p_{n_A})  \ ,
   \label{PnA3}
\end{aligned}
\ee
where
\be
   F_{n} (p, z ) \equiv  \int_0^1 \prod_{i=2}^{n-1} {d p_i
     \over p_i} \, \theta\left(p_2-{z \over p_2 p_3 \ldots
     p_{n-1}},  \, p_3 - p_2, \, \ldots, \, p - p_{n-1}\right) \ .
\ee
We now claim that
\be
  F_{n} (p, z) = {n^2 (n-1) \over (n!)^2} \ln^{n-2}
     \left({ p^{n-1} \over z} \right) \,
     \theta\left(p-z^{1/(n-1)}\right) \ ,
     \label{induct}
\ee
which we will prove by induction on $n$.  For $n = 3$,
Eq.~(\ref{induct}) can be verified by direct calculation. 
Suppose now that it holds for some $n$.  We can then calculate
$F_{n+1}(p, z)$ as follows. 
\be
\begin{aligned}
   F_{n+1} (p, z ) &=  \int_0^1 \prod_{i=2}^{n} {d p_i
     \over p_i} \, \theta\left(p_2-{z \over p_2 p_3 \ldots
     p_{n-1} p_n}, \, p_3 - p_2, \, \ldots, \, p_n
     - p_{n-1}, \, p - p_n \right) \\
   &= \int_0^1 {d p_n \over p_n} \theta(p - p_n) \\
   &\qquad \times \int_0^1 \prod_{i=2}^{n-1} {d p_i \over p_i} \,
     \theta\left(p_2-{z/p_n \over p_2 p_3 \ldots p_{n-1}},
     \, p_3 - p_2, \, \ldots, \, p_n -
     p_{n-1}  \right) \\
   &= \int_0^1 {d p_n \over p_n} \theta(p - p_n) \, F_n(p_n,
     z/p_n) \ .
\end{aligned}
\ee
Now using the induction hypothesis, we find
\be
  F_{n+1} (p, z ) =  {n^2 (n-1) \over (n!)^2} 
     \int_0^1 {d p_n \over p_n} \theta(p - p_n) \,
     \ln^{n-2} \left({p_n^{n-1} \over (z/p_n) }\right) \,
     \theta\left(p_n - \left({z \over p_n} \right)^{1/(n-1)}
     \right)\ .
\ee
The second $\theta$-function is equal to 1 if
\be
  p_n > \left({z \over p_n}\right)^{1/(n-1)} \Longleftrightarrow
     p_n^{n-1} > {z \over p_n} \Longleftrightarrow p_n^n > z
     \Longleftrightarrow p_n > z^{1/n} \ ,
\ee
so it can be rewritten as $\theta(p_n - z^{1/n})$.  The two
$\theta$-functions then provide upper and lower limits on the
integration, but the integral is nonzero only if the upper limit
is larger than the lower limit, i.e., if $p > z^{1/n}$.  Thus, 
\be
\begin{aligned}
  F_{n+1} (p, z ) &= {n^2 (n-1) \over (n!)^2}
     \int_{z^{1/n}}^p \, {d p_n \over p_n} \ln^{n-2} \left({p_n^n \over
     z}\right) \theta(p - z^{1/n}) \\
   &= {n^2 (n-1) \over (n!)^2} n^{n-2} \int_{z^{1/n}}^p \, {d p_n \over
     p_n} \ln^{n-2} \left({p_n \over z^{1/n}}\right)
     \theta\left(p - z^{1/n}\right) \ .
\end{aligned}
\ee
Now we can change the variable of integration to $\bar p \equiv
p_n/z^{1/n}$, so
\be
\begin{aligned}
   F_{n+1} (p, z)  &= {n^2 (n-1) \over (n!)^2} n^{n-2} 
     \int_1^{p/z^{1/n}} \, {d \bar p \over \bar p} \, \ln^{n-2}
     (\bar p) \,\theta\left(p - z^{1/n}\right) \\
   &=  {n^2 (n-1) \over (n!)^2} {n^{n-2} \over n-1} \ln^{n-1} \left({p
     \over z^{1/n}}\right)  \theta\left(p - z^{1/n}\right) \\
   &=  {n^2 (n-1) \over (n!)^2} {n^{n-2} \over (n-1) \, n^{n-1}}
     \ln^{n-1} \left({p^n \over z}\right) \theta\left(p -
     z^{1/n}\right) \\
   &= {(n+1)^2 n \over \bigl((n+1)!\bigr)^2}  \ln^{n-1} \left({p^n
     \over z}\right) \theta\left(p - z^{1/n}\right)
     \ ,
\end{aligned}
\ee
which verifies the induction hypothesis.

Finally, inserting Eq.~(\ref{induct}) into Eq.~(\ref{PnA3}), we
find the result that was stated in the text as Eq.~(\ref{PnA}).

\end{document}